\documentclass[preprint]{revtex4}
\usepackage{epsfig}
\usepackage{amsmath}

\begin{document}

\title{Semi-transparent brane-worlds}
\author{Zolt\'{a}n Keresztes$^{1}$, Ibolya K\'{e}p\'{\i}r\'{o}$^{2}$, and 
L\'{a}szl\'{o} \'{A}. Gergely$^{1}$}
\affiliation{$^{1}$Departments of Theoretical and Experimental Physics, University of
Szeged, 6720 Szeged, D\'{o}m t\'{e}r 9, Hungary\\
$^{2}$Blackett Laboratory, Imperial College, Prince Consort Road, London SW7
2BW, UK}

\begin{abstract}
We study the evolution of a closed Friedmann brane perturbed by the Hawking
radiation escaping a bulk black hole. The semi-transparent brane absorbes
some of the infalling radiation, the rest being transmitted across the brane
to the other bulk region. We characterize the cosmological evolution in
terms of the transmission rate $\varepsilon $. For small values of $
\varepsilon $ a critical-like behaviour could be observed, when the
acceleration due to radiation pressure and the deceleration induced by the
increasing self-gravity of the brane roughly compensate each other, and
cosmological evolution is approximately the same as without radiation.
Lighter (heavier) branes than those with the critical energy density will
recollapse slower (faster). This feature is obstructed at high values of $
\varepsilon $, where the overall effect of the radiation is to speed-up the
recollapse. We determine the maximal value of the transmission rate for
which the critical-like behaviour is observed. We also study the effect of
transmission on the evolution of different source terms of the Friedmann
equation. We conclude that among all semi-transparent branes the slowest
recollapse occurs for light branes with total absorption.
\end{abstract}

\email{zkeresztes@titan.physx.u-szeged.hu , ibolya.kepiro@imperial.ac.uk ,
gergely@physx.u-szeged.hu}
\date{\today }
\maketitle


\section{Introduction}

Enlarging the number of dimensions of space-time is a well established idea
dating back to Kaluza and Klein. Originally such attempts were ment to
incorporate the non-gravitational interactions into a geometric framework,
and this could be partially achieved by adding non-compact spatial
dimensions to the four-dimensional space-time.

A recent development in the subject was the suggestion to enrich the
space-time by one non-compact dimension \cite{RS}. Starting from the
assumption that gravitation is an effectively five-dimensional interaction,
while the particles and fields of the standard model are effectively
four-dimensional, we arrive to a novel theory of gravitation, known as
brane-world gravity. In this theory our four-dimensional space-time is a
hypersurface (the brane), embedded in the five-dimensional space-time (the
bulk). In a generic setup both the brane and the bulk can be curved. The
brane has a positive tension $\lambda $, which together with the
5-dimensional coupling constant $\widetilde{\kappa }^{2}$ defines the
four-dimensional coupling constant $\kappa ^{2}$ as 
\begin{equation}
6\kappa ^{2}=\widetilde{\kappa }^{4}\lambda \,.
\end{equation}
The matter on the brane is induced by the embedding of the brane into the
bulk. The Lanczos-Sen-Darmois-Israel junction conditions \cite{Lanczos}-\cite
{Israel} relate the brane energy-momentum tensor $T_{ab}$ (together with the
brane tension) to the jump across the brane in the extrinsic curvature $
K_{ab}$. In the simplest case, when the setup is symmetric; the extrinsic
curvature on one side is exactly opposite to the extrinsic curvature on the
other side. Thus the jump in the extrinsic curvature is simply twice the
value of the extrinsic curvature on one side of the brane. However
asymmetric embeddings were also considered \cite{Kraus}-\cite{Stoica}. Then
the theory acquires interesting new features, like late-time acceleration in
cosmological branes.

The bulk contains at least a negative cosmological constant $\widetilde{
\Lambda }.$ In the simplest case $\widetilde{\Lambda }$ can be fine-tuned to
the brane tension such that the effective cosmological constant on the brane
vanishes (Randall-Sundrum gauge, \cite{RS}). In a more generic setup $
\widetilde{\Lambda }$ and $\lambda $ can combine in such a way, that the
brane remains with a tiny cosmological constant $\Lambda $. In the most
generic case, when we allow for both non-standard matter in the bulk (like a
scalar feld, possibly taking over the role of the bulk cosmological
constant, or a radiation of quantum origin) and for asymmetric embedding,
the brane cosmological constant is given by the relation 
\begin{equation}
2\Lambda =\kappa ^{2}\lambda +\widetilde{\kappa }^{2}\overline{\widetilde{
\Lambda }}-\frac{\overline{L}}{2}-\widetilde{\kappa }^{2}\overline{ 
\widetilde{T}_{cd}n^{c}n^{d}}\ ,\   \label{Lambda}
\end{equation}
where $\widetilde{T}_{cd}$ represents the energy-momentum tensor of the
non-standard model bulk fields, $\ n^{a}$ is the normal to the brane and $
\overline{L}$ is the trace of 
\begin{equation}
\overline{L}_{ab}=\overline{K}_{ab}\overline{K}-\overline{K}_{ac}\overline{K}
_{b}^{c}-\frac{g_{ab}}{2}\left( \overline{K}^{2}-\overline{K}_{ab}\overline{
K }^{ab}\right) \ .
\end{equation}
.An overbar denotes the average of a quantity, taken over the left and right
sides of the brane. Therefore for a symmetric embedding $\overline{K}
_{ab}=0= \overline{L}$. We remark that for either $\overline{L}\neq 0$ or $
\overline{ \widetilde{T}_{cd}n^{c}n^{d}}\neq 0$, the quantity $\Lambda $
defined by Eq. (\ref{Lambda}) can fail to be a constant \cite{Decomp}.

On the brane, gravitational dynamics is modified as compared to general
relativity. It is governed by a modified Einstein equation, derived in \cite
{Decomp} as a generalization of the covariant approach of \cite{SMS}: 
\begin{equation}
G_{ab}=-\Lambda g_{ab}+\kappa ^{2}T_{ab}+\widetilde{\kappa }^{4}S_{ab}-
\overline{\mathcal{E}}_{ab}+\overline{L}_{ab}^{TF}+\overline{\mathcal{P}}
_{ab}\ ,  \label{modEgen}
\end{equation}
where $S_{ab}$ represents a quadratic expression in $T_{ab}$: 
\begin{equation}
S_{ab}=\frac{1}{4}\Biggl[-T_{ac}^{\ }T_{b}^{c}+\frac{1}{3}TT_{ab}-\frac{
g_{ab}}{2}\left( -T_{cd}^{\ }T^{cd}+\frac{1}{3}T^{2}\right) \Biggr]\ ,
\label{S}
\end{equation}
and the source term $\overline{\mathcal{E}}_{ab}$ is the average of the
electric part of the bulk Weyl tensor $\widetilde{C}_{abcd}$, defined as 
\begin{equation}
\mathcal{E}_{ac}=\widetilde{C}_{abcd}n^{b}n^{d}\ .  \label{calE}
\end{equation}
The last source term is from the pull-back of the bulk sources to the brane 
\begin{equation*}
\overline{\mathcal{P}}_{ab}=\frac{2\widetilde{\kappa }^{2}}{3}\overline{
{}\left( \widetilde{T}_{cd}g_{a}^{c}{}g_{b}^{d}\right) ^{TF}}\ ,
\end{equation*}
and $TF$ stands for trace-free. The source terms $T_{ab}$ and $S_{ab}$ are
local, the latter modifying gravitational dynamics at high energies. By
contrast, at low energies the effect of $S_{ab}$ is negligible as compared
to $T_{ab}$. Another feature is that the evolution of the source terms $-
\overline{\mathcal{E}}_{ab}+\overline{L}_{ab}^{TF}+\overline{\mathcal{P}}
_{ab}$ is non-local, as it depends on the bulk degrees of freedom. In order
to close the system, the modified Einstein equation (\ref{modEgen}) has to
be supplemented by the Codazzi equation and the twice contracted Gauss
equation \cite{Decomp}, as well as by evolution equations for the bulk
fields.

When branes with cosmological symmetry are embedded into a Schwarzschild -
anti de Sitter bulk (SAdS5), standard cosmological evolution emerges at low
energies. By adding to the model an asymmetry in the bulk cosmological
constant, a slight late-time acceleration appears (see \cite{Induced} and
references therein). Cosmological branes embedded in a bulk containing
radiation escaping from the brane were studied in \cite{ChKN}, \cite{LSR}, 
\cite{RadBrane}, \cite{VernonJennings} and \cite{Langlois}.

Other models, with Hawking radiation escaping from a bulk black hole, were
also studied. These include branes with $k=0$ \cite{Jennings} and closed
brane-worlds \cite{GK}. In the latter model $k=1$ was chosen because the
expression for the energy density of the Hawking radiation was originally
derived for closed universes \cite{EHM}- \cite{GCL}). Other essential
features of the model discussed in \cite{GK} were the existence of only one
bulk black hole, the radiation of which was completely absorbed on the
brane. (By contrast, in the model \cite{Jennings}, there were two bulk black
holes and the radiation was completely transmitted.) The final conclusion of 
\cite{GK} was that the recollapsing fate of the $k=1$ universe could not be
avoided neither by the asymmetry due to the location of the single bulk
black hole, nor by the absorbed bulk radiation. Indeed, the radiation
contributed by two competing effects: a radiation pressure, accelerating the
brane and an increase in the self-gravity of the brane (due to absorption),
which contributed to a faster recollapse. For the case studied of total
absorption on \ the brane a critical behaviour was observed for a certain
value of the initial brane energy density, when these two effects roughly
cancel each other, and cosmological evolution proceeds (roughly) as in the
absence of radiation.

In this paper we introduce a new degree of freedom into the model of \cite
{GK}. Specifically, we allow for the \textit{transmission} across the brane.
Thus the brane we study here is semi-transparent, in contrast with the
completely opaque brane discussed in \cite{GK}. The rate of transmission is
given by a parameter $\varepsilon $, zero for total absorption, and one for
total transmission. We continue to neglect the reflection because there is
no exact solution with cosmological constant describing a crossflow of
radiation streams (although such a solution is known in four dimensions, in
the absence of a cosmological constant \cite{GergelyND}). According to these
assumptions, the five-dimensional Vaidya-anti de Sitter (VAdS5) spacetime
describes both bulk regions.

In Section 2 we develop the mathematics of our model. The radiation pressure
accelerating the brane is obviously decreasing with increasing $\varepsilon $
. The absorbed radiation, and the increase of self-gravity due to this is
again decreasing with increasing $\varepsilon $. The question comes, how
these two effects may change the cosmological evolution. Having set the
relevant equations in dimensionless variables, we carry out a numerical
analysis of the cosmological evolution in Section 3 and we present graphical
solutions. We summarize our results on how the semi-transparency of the
brane affects the conclusions traced in \cite{GK} in the Concluding Remarks.

\section{Dynamics of the semi-transparent brane}

When the brane obeys cosmological symmetries, for $k=1$ it is described by
the line element 
\begin{equation}
ds^{2}=-dt^{2}+a^{2}\left( t\right) \left[ \frac{dr^{2}}{1+r^{2}}
+r^{2}\left( d\theta ^{2}+\sin ^{2}\theta ~d\varphi ^{2}\right) \right] ~.
\end{equation}
Here $a$ denotes the scale factor. The modified Einstein equations (\ref
{modEgen}) decouple into a generalized Friedmann and a generalized
Raychaudhuri equation \cite{Decomp}. As shown in \cite{Decomp}, the
difference of the Codazzi equations, taken on the two sides of the brane
gives the energy-balance equation. The last relevant equation is the energy
emission of the bulk black hole.

We would like to write these equations in terms of dimensionless variables
(these will be distinguished by a hat from the corresponding dimensional
variables). Following \cite{GK}, we define: 
\begin{eqnarray}
\widehat{t} &=&Ct\text{ },\text{\ }\widehat{a}=Ca\text{ },\text{\ }\widehat{
H }=\frac{H}{C}\text{ },  \notag \\
\text{ }\widehat{\rho } &=&\frac{\rho }{\lambda }\text{ },\text{ }\widehat{p}
=\frac{p}{\lambda }\text{ },\text{ }\widehat{\psi }=\frac{\psi }{C\lambda } 
\text{ },  \notag \\
\text{ \ }\widehat{\Delta m} &=&C^{2}\Delta m\text{ \ },\text{ }\widehat{ 
\overline{m}}=C^{2}\overline{m}\text{ }.
\end{eqnarray}
Here $C=\kappa \sqrt{\lambda }$, and units $c=\hbar =1$ were chosen. $H$ is
the Hubble parameter, $\rho $ is the brane energy density, $p$ the pressure
of the matter on the brane. The energy density of the radiation escaping the
bulk black hole is denoted as $\psi $. The quantity $\Delta m$ represents
the difference of the mass \textit{functions} of the VAdS5 metric and $
\overline{m}$ is their average.

The brane evolves cf. the energy balance, Friedmann- and Rachaudhuri
equations \cite{Decomp}. The energy balance equation, written in
dimensionless variables is 
\begin{equation}
\widehat{\rho }^{^{\prime }}+3\widehat{H}\left( \widehat{\rho }+\widehat{p}
\right) =(1-\varepsilon )\widehat{\psi }\ ,  \label{energy balance}
\end{equation}
the Raychaudhuri equation reads 
\begin{eqnarray}
\widehat{H}^{^{\prime }} &=&-\widehat{H}^{2}-\frac{1}{6}\widehat{\rho }
\left( 1+2\widehat{\rho }\right) -\frac{1}{2}\widehat{p}\left( 1+\widehat{
\rho }\right) -\frac{2\widehat{\overline{m}}}{\widehat{a}^{4}}  \notag \\
&&+\frac{9\left( \widehat{p}-1\right) \left( \widehat{\Delta m}\right) ^{2}}{
2\widehat{a}^{8}\left( \widehat{\rho }+1\right) ^{3}}-\frac{(1+\varepsilon )
\widehat{\psi }}{\sqrt{6}}+\frac{3(1-\varepsilon )\text{\ }\widehat{\psi }
\widehat{\Delta m}}{\sqrt{6}\widehat{a}^{4}\left( \widehat{\rho }+1\right)
^{2}}\ ,  \label{Raychaudhuri}
\end{eqnarray}
and the Friedmann equation is 
\begin{equation}
\widehat{H}^{2}=-\frac{1}{\widehat{a}^{2}}+\frac{\widehat{\rho }}{3}\left( 1+
\frac{\widehat{\rho }}{2}\right) +\frac{2\widehat{\overline{m}}}{\widehat{a}
^{4}}+\frac{3\left( \widehat{\Delta m}\right) ^{2}}{2\widehat{a}^{8}\left( 
\widehat{\rho }+1\right) ^{2}}\ .  \label{Friedmann}
\end{equation}
The energy density of the radiation escaping from the black hole, evaluated
at the left side of the brane becomes \cite{GK}: 
\begin{equation}
\widehat{\psi }=\frac{5\kappa ^{4}\lambda \zeta \left( 5\right) }{24\pi ^{9}
\widehat{a}^{4}\widehat{m}_{L}\left\{ \widehat{H}+\left( \widehat{\rho }
+1\right) /\sqrt{6}+3\widehat{\Delta m}/\left[ \sqrt{6}\widehat{a}^{4}\left( 
\widehat{\rho }+1\right) \right] \right\} }\text{ },  \label{psihat}
\end{equation}
where $\zeta $ is the Riemann-zeta function. Here $\widehat{m}_{L}=\widehat{
\overline{m}}-\widehat{\Delta m}/2$ is the value of the mass function near
the left side of the brane. It contains both the mass of the black hole and
the energy of the Hawking radiation. From (\ref{psihat}) and the formulae
(90), (98) and (99) of\ \cite{Decomp} employed for our model; by using the
relations giving the embedding of the brane, we derive the evolutions of $
\widehat{\overline{m}}$ and $\widehat{\Delta m}$: 
\begin{eqnarray}
\widehat{\overline{m}}^{^{\prime }} &=&-\frac{\widehat{a}^{4}\widehat{\psi }
}{\sqrt{6}}\left\{ \left( 1+\varepsilon \right) \left[ \widehat{H}+\frac{3
\widehat{\Delta m}}{\sqrt{6}\widehat{a}^{4}\left( \widehat{\rho }+1\right) }
\right] +\left( 1-\varepsilon \right) \frac{\widehat{\rho }+1}{\sqrt{6}}
\right\} \text{ },  \label{mbar} \\
\widehat{\Delta m}^{^{\prime }} &=&\frac{2\widehat{a}^{4}\widehat{\psi }}{
\sqrt{6}}\left\{ \left( 1-\varepsilon \right) \left[ \widehat{H}+\frac{3
\widehat{\Delta m}}{\sqrt{6}\widehat{a}^{4}\left( \widehat{\rho }+1\right) }
\right] +\left( 1+\varepsilon \right) \frac{\widehat{\rho }+1}{\sqrt{6}}
\right\} \text{ }.  \label{deltam}
\end{eqnarray}
Eqs. (\ref{energy balance}), (\ref{Raychaudhuri}), (\ref{Friedmann}), (\ref
{psihat}), (\ref{mbar}) and (\ref{deltam}) give a closed, but complicated
system of coupled differential equations, from which further information can
be extracted only by a numerical study. Eqs. (\ref{mbar}) and (\ref{deltam})
give the evolutions of $\widehat{m}_{L}=\widehat{\overline{m}}-\widehat{
\Delta m}/2$ and $\widehat{m}_{R}=\widehat{\overline{m}}+\widehat{\Delta m}/2
$. The evolution of $\widehat{m}_{L}$ separates from the system and it is
integrable. The result is given by the formula (38) of \cite{GK}, which is
independent of the rate of transmission. This can happen because the
evaporation of the bulk black hole does not depend on the transmission rate
across the brane.

\section{Cosmological evolution of the semi-transparent brane}

The partial transmission across the brane induces new features in the model.
First, the accumulated energy from the absorbed Hawking radiation on the
brane will be smaller than in the case of total absorption. Second, the
transmitted radiation does not contribute to the radiation pressure on the
brane, which also becomes smaller. We also note, that without transmission
and with only one black hole in the bulk, one of the bulk regions was
SAdS5 in \cite{GK}, as opposed to the present case,
when both regions are VAdS5. 
\begin{figure}[th]
\centering\includegraphics[width=0.55\linewidth]{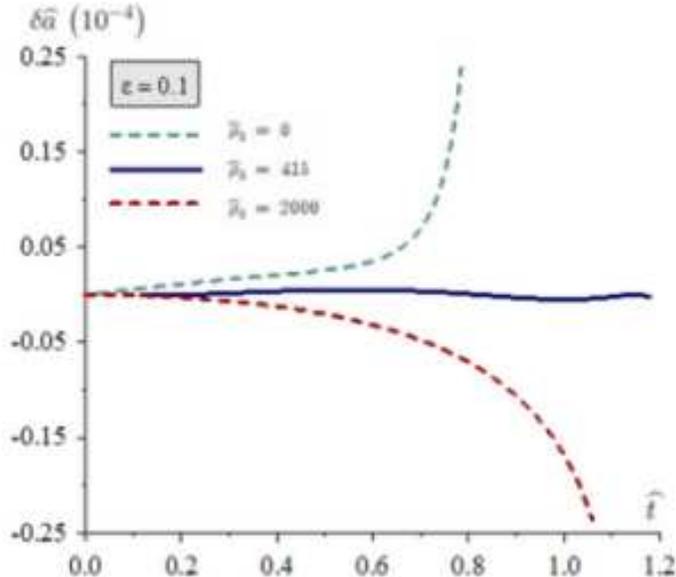}
\caption{Time evolution of the difference $\protect\delta \widehat{a}$
between the (dimensionless) scale factors in the radiating and non-radiating
cases. For the depicted transmission rate of $\protect\varepsilon =0.1$ the
critical-like behaviour $\protect\delta \widehat{a}\approx 0$ is observed at 
$\widehat{\protect\rho }_{0}^{crit}=415$ (solid curve). This critical value
of the initial energy density is smaller than for $\protect\varepsilon =0$
(then $\widehat{\protect\rho }_{0}^{crit}=520$ was observed, see 
\protect\cite{GK}). For small initial energy densities (in the extreme case $
\widehat{\protect\rho }_{0}=0$, upper dashed curve), the pressure of the
Hawking radiation dominates. For high initial energy densities (like $
\widehat{\protect\rho }_{0}=2000$, lower dashed curve) the increase of
self-gravity due to absorption overtakes the radiation pressure.}
\label{Fig1}
\end{figure}
\begin{figure}[th]
\centering\includegraphics[width=0.55\linewidth]{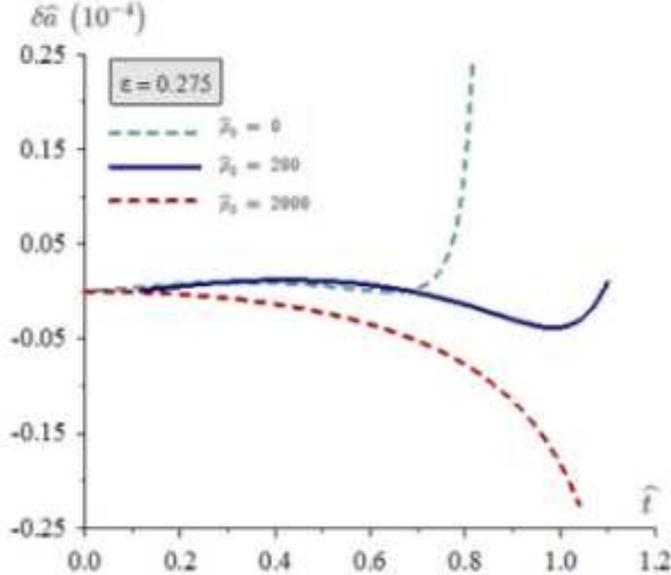}
\caption{As in Fig. \protect\ref{Fig1}, but for a higher transmission rate $
\protect\varepsilon =0.275$. The \textit{critical} brane initial energy
density further decreases to $\widehat{\protect\rho }_{0}^{crit}=200$.
Moreover, the amplitude of the sinusoidal evolution of $\protect\delta 
\widehat{a}$ on the critical curve increases, damping the critical-like
behaviour as compared to smaller absorption rates.}
\label{Fig2}
\end{figure}
\begin{figure}[th]
\centering{$(a)$\includegraphics[width=0.55\linewidth]{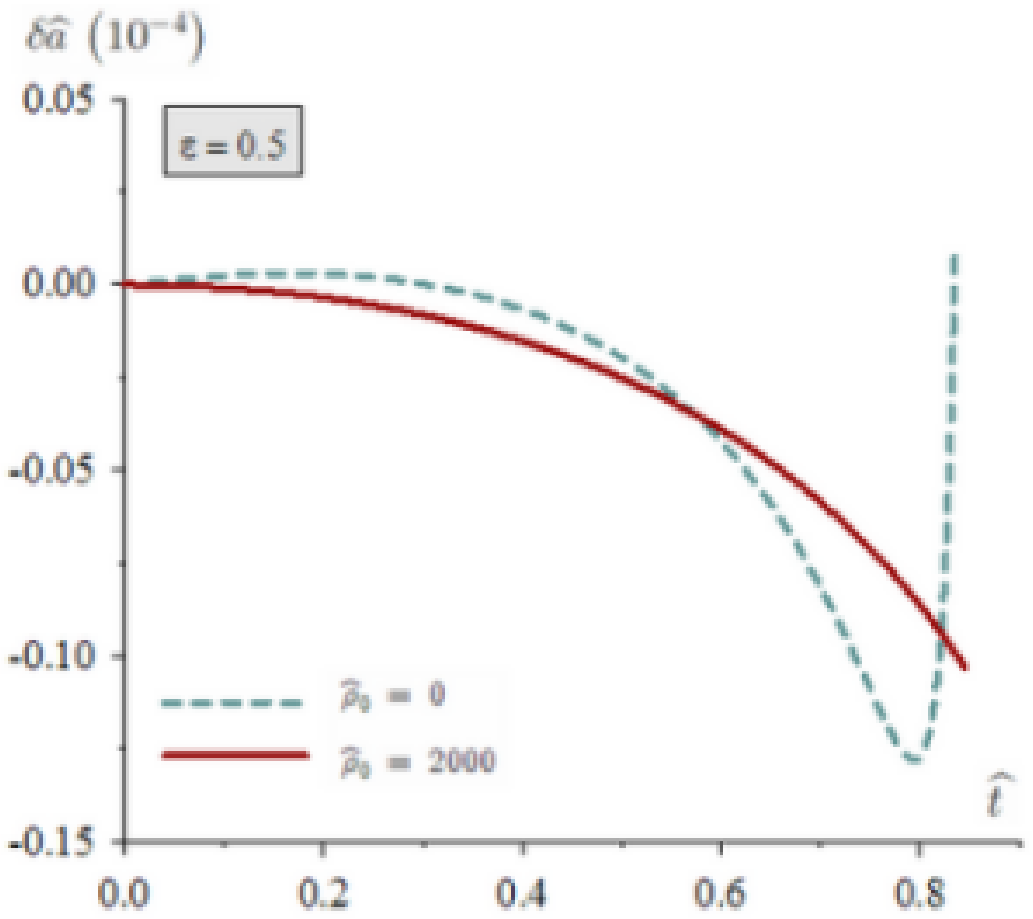}} 
\centering
{$(b)$\includegraphics[width=0.55\linewidth]{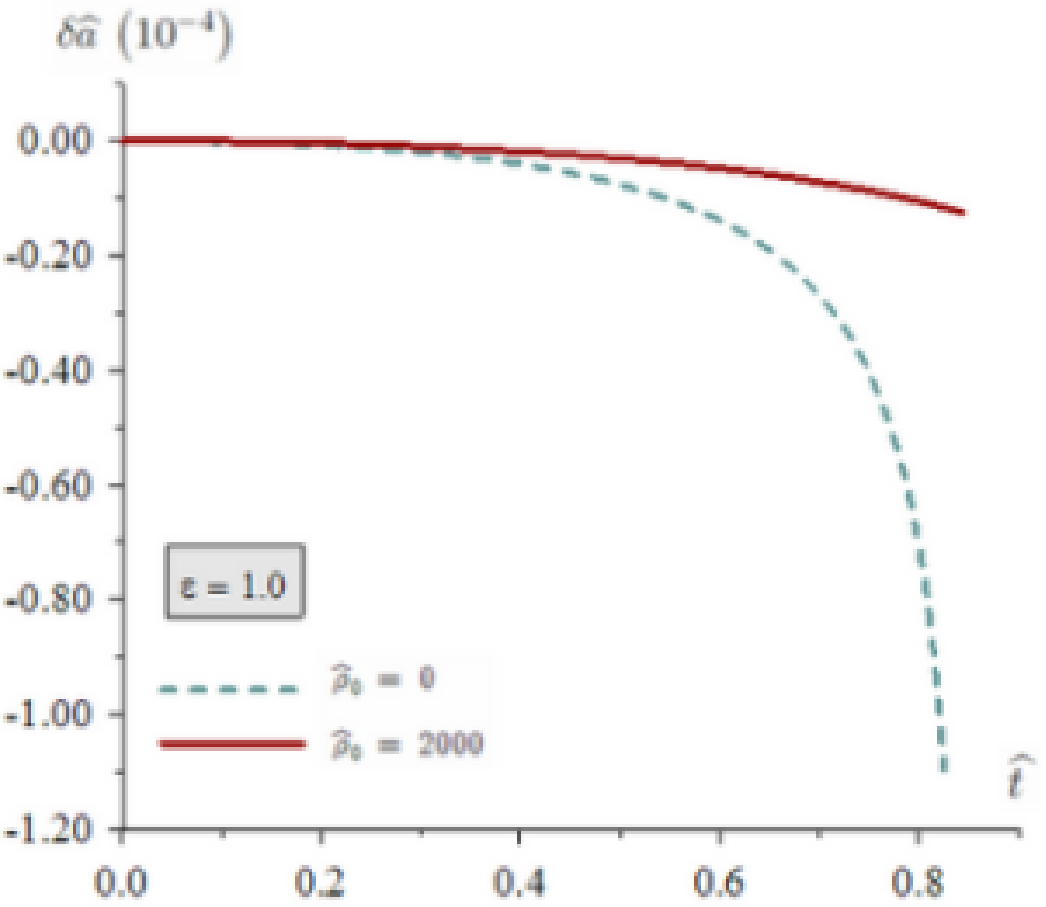}}
\caption{For even higher transmission rates, there is no critical-like brane
evolution at all. Therefore the time evolution of the difference $\protect
\delta \widehat{a}$ between the (dimensionless) scale factors in the
radiating and non-radiating cases is depicted only for the initial values $
\widehat{\protect\rho }_{0}=0$ (solid curve) and $\widehat{\protect\rho }
_{0}=2000$ (dashed curve). For any value of the initial energy density the
net contribution of the radiation is towards a faster recollapse. The cases
shown are (a) $\protect\varepsilon =0.5$ (b) total transmision $\protect
\varepsilon =1$.}
\label{Fig3}
\end{figure}
\begin{figure}[th]
\centering\includegraphics[width=0.55\linewidth]{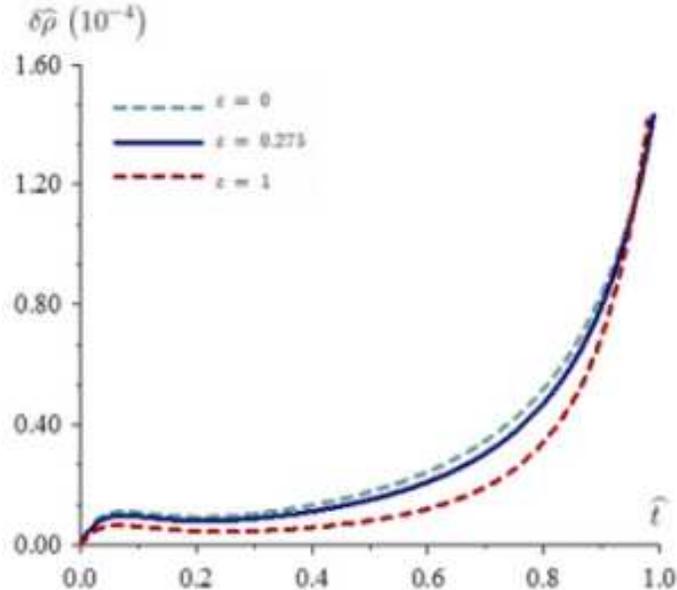}
\caption{Time evolution of the perturbation caused by the Hawking radiation
in the brane energy density, for initial brane energy density $\widehat{
\protect\rho }_{0}=300$ and transmission rates $\protect\varepsilon =0$, $
0.275$ and $1$.}
\label{Fig4}
\end{figure}
\begin{figure}[th]
\centering{\includegraphics[width=0.55\linewidth]{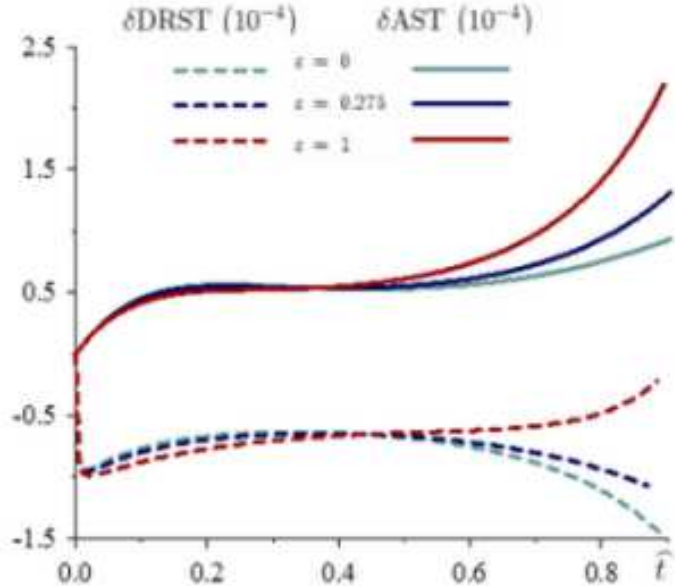}}
\caption{Time evolution of the perturbations caused by the Hawking radiation
in the source terms of the Friedmann equation. All solid lines represent
perturbations in the \textit{asymmetry source term,} for transmission rates $
\protect\varepsilon =0$, $0.275$ and $1$. All dotted lines represent
perturbations in the \textit{dark radiation source term}, for the same
transmission rates. The plot is for the\ initial brane energy density $
\widehat{\protect\rho }_{0}=520$ (the critical value of the brane initial
energy density for opaque branes).}
\label{Fig5}
\end{figure}
\begin{figure}[th]
\centering{$(a)$\includegraphics[width=0.55\linewidth]{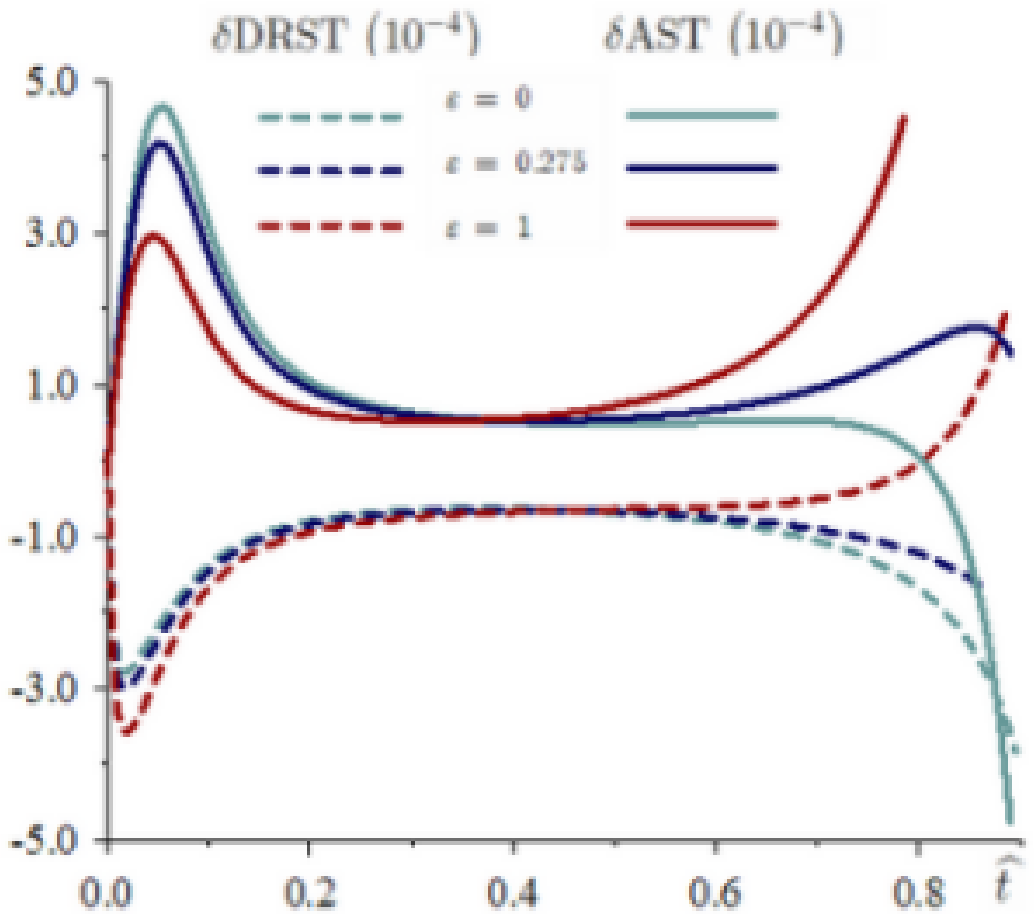}} 
\centering
{$(b)$\includegraphics[width=0.55\linewidth]{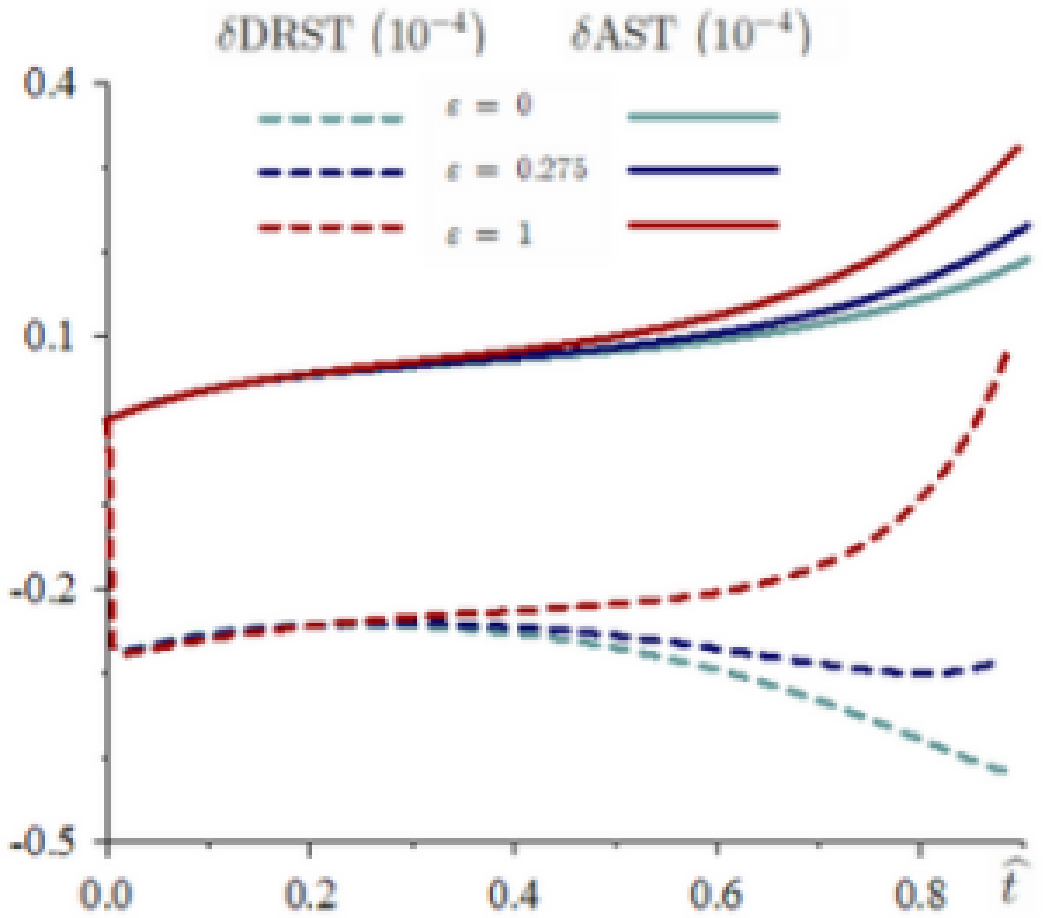}}
\caption{As in Fig. \protect\ref{Fig5}, but for light and heavy branes,
respectively. The plots are for the\ initial energy densities (a) $\widehat{
\protect\rho }_{0}=100$ and (b) $2000$.}
\label{Fig6}
\end{figure}

For the completely opaque branes discussed in \cite{GK}, both the
acceleration from the radiation pressure and the deceleration from the
increase in the self-gravity of the brane were small perturbations of order $
10^{-4}$, which roughly cancelled each other for the critical initial energy
density. As the semi-transparency of the branes further diminishes both
effects of the radiation, the perturbations will be even smaller here,
although of the same order of magnitude. We again plot only the \textit{
differences} of the physical quantities characterizing cosmological
evolution, taken in the radiating and non-radiating cases.

As in \cite{GK}, we take a radiation-dominated brane (with $\widehat{p}= 
\widehat{\rho }/3$) and start the evolution of the brane at the apparent
horizon of the bulk black hole. For more details on the chosen initial data
see \cite{GK}.

Figures \ref{Fig1}-\ref{Fig3} show the evolution of the differences in the
scale factor when the radiation is switched on and off, for increasing
transmission rates. Their sequence shows that the critical-like behaviour is
seriously deteriorated by increasing transmission across the brane. For tiny
values of $\varepsilon ~$we have found that the value of the critical energy
density\ decreases with increasing transmission rate. Also the
sinusoidal-like pattern of the critical curve is more accentuated, as the
transmission increases (Fig. \ref{Fig1}-\ref{Fig2}). The highest value of
the transmission rate, for which such a critical-like behaviour could be
observed is for $\varepsilon =0.275$ (Fig. \ref{Fig2}). For $\varepsilon
>0.275$, there is no value of the initial energy density for which the scale
factor is larger in the presence of the radiation than in the non-radiating
case. This holds true during the whole cosmological evolution. Therefore no
critical-like evolution of the perturbation in the scale factor can be
observed for these higher values of the transmission (Fig \ref{Fig3}).

For total transmission, there is no radiation pressure on the brane and no
increase in its self-gravity due to absorption at all. In this case, the
Hawking radiation appears only in the Raychaudhuri equation through the term 
$-(1+\varepsilon )\widehat{\psi }/\sqrt{6}$. This term contributes toward
deceleration, thus driving the universe towards a faster recollapse. In
consequence the scale factor is smaller than in the non-radiating case (Fig. 
\ref{Fig3}b).

Fig. \ref{Fig4} shows the time evolution of the difference between the brane
energy densities in the radiating and non-radiating cases, plotted for $
\widehat{\rho }_{0}=300$ and various transmission rates. We see that the
evolution of the brane energy density is not very much affected by the
transmission, rather it is just slightly rescaled.

Fig. \ref{Fig5}, shows the time evolution of the perturbation caused by the
bulk Hawking radiation in two of the source terms of the Friedmann equation.
These are the asymmetry and the dark radiation terms (defined as in \cite{GK}
). The two terms are roughly equal, but of opposite sign for a \textit{
critical} initial energy density $\widehat{\rho }_{0}^{crit}=520$,
corresponding to the $\varepsilon =0$ case, with the exceptions of the very
early and very late stages of the evolution of the Friedmann brane \cite{GK}
. By increasing the transmission, the evolutions will change drastically the
late-time behaviour (Fig. \ref{Fig5}). Furthermore, the changes induced at
late times by transmission are significant, regardless of the particular
value of the initial energy density (Fig. \ref{Fig6}a is for light branes,
while Fig. \ref{Fig6}b for heavy branes).

At early times the transmission induces changes only for light branes,
characterized by low initial energy density values (see \ Fig. \ref{Fig6}a).
We also remark that in the early stages of evolution of light branes a
higher value of the transmission rate decreases the magnitude of the
asymmetry source term, however the magnitude of the dark radiation source
term is increased. This tendency is reversed in the late stages of evolution.

\section{Concluding remarks}

In this paper we have considered a highly asymmetric, closed brane-world
model, with only one black hole in the bulk. The black hole is emitting
Hawking radiation and this is partially transmitted through the brane to the
other bulk region. The modifications to standard cosmological evolution
caused by both the asymmetric setup and the radiation in the bulk represent
small perturbations. By varying the transmission rate from total absorption
(opaque brane) to total transmission we have seen, how the rate of
transmission affects these perturbations. We have determined numerically the
value $\varepsilon =0.275$ of the transmission rate for wich the
critical-like behaviour discussed in \cite{GK} disappears. For all branes
with a high degree of opacity the critical behaviour can be found for
certain value of the initial energy density. The bigger the transmission
rate, the lower this energy density and the lighter the brane with critical
behaviour. For $\varepsilon =0.275$ the critical initial energy density on
the brane would be zero.

We have also studied the evolution of the dark radiation and asymmetry
source terms of the Friedmann equation. The evolution of these sorce terms
was also discussed in \cite{Jennings}, but in a different setup ($k=0$, $
\varepsilon =1$, two black holes in the bulk and asymmetry in the
cosmological constant). In that model the dark radiation term decreased,
while the asymmetry term increased due to the black hole radiation. Our
investigations lead to the same conclusion, however they also show the
modulation of these effects due to the transmission rate. The suppression of
the dark radiation term is more accentuated for high $\varepsilon $ at early
times and for small $\varepsilon $ at late times. By contrast the asymmetry
term increases more significantly for small $\varepsilon $ at early times
and for high $\varepsilon $ at late times.

As a generic rule we have found that the semi-transparent brane-worlds
recollapse faster when the transmission rate is high. Thus the opaque branes
discussed in \cite{GK} live the longest while the fastest recollapse occurs
for total transmission.

\begin{acknowledgments}
This work was supported by OTKA grants no. T046939 and TS044665. L\'{A}G
wishes to thank the support of the J\'{a}nos Bolyai Scholarship of the
Hungarian Academy of Sciences.
\end{acknowledgments}

\end{document}